\documentclass[amsmath,amssymb,aps,reprint]{revtex4-2}

\usepackage{graphicx} 
\usepackage{newtxtext,newtxmath}
\usepackage{physics}
\usepackage{svg}
\usepackage{bm} 
\usepackage{hyperref}
\hypersetup{colorlinks,linkcolor={blue},citecolor={blue},urlcolor={red}}

\begin{document}

\title{
\textbf{Octahedral Rotation Induced, Antiferroelectric-like Double Hysteresis in Strained Perovskites}
}

\author{
Seongjoo Jung and Turan Birol$^{*}$
}

\affiliation{Department of Chemical Engineering and Materials Science, University of Minnesota, Minneapolis, Minnesota 55455, USA}

\begin{abstract}
Antiferroelectrics, which host both polar and antipolar order parameters, are characterized by the double hysteresis loops which are advantageous for various applications such as high-density energy storage. 
In this study, we investigate the coupling between oxygen octahedral rotations and polarization in well-known perovskites, with a focus on SrTiO$_3$. 
Using first-principles calculations and symmetry-adapted Landau-Ginzburg-Devonshire theory, we construct an energy landscape to analyze how this coupling shapes polarization-voltage hysteresis behavior. 
We show that tuning the relative strength of polar and rotational instabilities by exploiting epitaxial strain and layering leads to nontrivial hysteresis behavior. Consequently, the rotation coupling with polarization leads to an expanded search space of materials exhibiting antiferroelectric-like double hysteresis.
\end{abstract}

\maketitle

Electric field responses of insulators, including piezoelectric, pyroelectric, ferroelectric and electrooptical responses \cite{scott2007applications,martin2016thin,safaei2019review, zhang2021recent}, are widely utilized in technological applications. While most insulators are dielectrics with linear polarization-electric field ($P$-$\mathcal{E}$) relationships, there also exist materials with more complex responses. Ferroelectrics (FEs) are insulators that can exhibit spontaneous polarization, characterized by hysteresis loops at the center of their $P$-$\mathcal{E}$ curves. FEs necessarily lack inversion symmetry, and can be used in applications such as data storage  \cite{buck1952ferroelectrics}, low power transistors \cite{salahuddin2008use}, in addition to exhibiting the aforementioned piezoelectric and pyroelectric effects.

Antiferroelectrics (AFEs) are another class of insulators distinguished by their polarization response, characterized by antiparallel dipole moments in their low-temperature ground state. Under an electric field, they transition to a polar state with parallel dipoles, producing a characteristic double hysteresis loop in the $P$-$\mathcal{E}$ curve. This switchable behavior enables a wide range of applications, including a strong electrocaloric effect, electromechanical responses, thermal switching, and photovoltaic effects \cite{si2024antiferroelectric,geng2015giant,yao2023ferrielectricity,liu2023low,perez2016above}. Furthermore, their double hysteresis makes AFEs ideal for high-energy-density capacitors \cite{zhao2017lead,kim2023ultrahigh}. Despite the discovery of AFE PbZrO$_3$ decades ago \cite{shirane1951dielectric,randall2021antiferroelectrics}, most applications still rely on PbZrO$_3$-based systems, as significantly fewer AFEs have been identified compared to FEs \cite{rabe2013antiferroelectricity}.

A commonly used model of antiferroelectricity is the Kittel model \cite{kittel1951theory}, which was further elaborated by Cross \cite{cross1967antiferroelectric} and Okada \cite{okada1969phenomenological}. These models employ a free energy expression that incorporates the coupling between a polar order parameter (polarization), $P$, and an antipolar order parameter, $Q$. The hysteresis observed at the phase boundary between FE and AFE phases arises from large variation and discontinuity in $\dd Q/ \dd P$, which subsequently leads to a sign change and discontinuity in $\dd \mathcal{E}/ \dd P$. Recent findings challenge the traditional view, suggesting that antiferroelectricity does not necessarily require the antipolar orientation to be parallel to the polarization or the applied field \cite{xue2015composition}. Furthermore, AFE-like hysteresis has been reported in BaTiO$_3$-SrTiO$_3$ superlattice structures, attributed to multi-domain polar vortices \cite{aramberri2022ferroelectric}, broadening the conventional understanding; and there is also evidence of AFE behavior in [BaZrO$_3$]$_7$-[SrTiO$_3$]$_7$ superlattices \cite{christen2003ferroelectric}, where there are open questions about the underlying mechanism. This highlights the need for further exploration of novel mechanisms responsible for the observed nontrivial hysteresis behavior.

This study investigates the potential for discovering AFE-like double hysteresis behavior resulting from the coupling of polarization with a structural order parameter that is not typically classified as antipolar \cite{toledano2016theory}. Specifically, we aim to illustrate this phenomenon in the context of the most common structural distortion in perovskites: octahedral rotations \cite{woodward1997octahedral}. SrTiO$_3$, a material extensively examined both experimentally and theoretically, is an ideal candidate for this demonstration. As a quantum paraelectric material \cite{muller1979srti, zhong1996effect}, SrTiO$_3$ exhibits phonon instabilities at the $\Gamma$-, R-, and M-points of the Brillouin zone in the cubic phase in density functional theory (DFT) calculations \cite{himmetoglu2014first}. While the polar instability is suppressed by quantum fluctuations \cite{zhong1996effect}, DFT predicts its ground state to host a polarization along with a$^0$a$^0$c$^-$ octahedral rotations (also called the antiferrodistortive (AFD) distortions) around the polar axis. The polar and AFD modes soften by compressive (out-of-plane) and tensile (in-plane) biaxial strain \cite{antons2005tunability, erba2013piezoelectricity, angsten2018epitaxial}, and the softening of the polar mode has been experimentally shown for SrTiO$_3$ under both tensile and compressive biaxial strain \cite{haeni2004room, wordenweber2007ferroelectric}.

In this study, we explore the coupling between the polarization and AFD rotation in epitaxially strained SrTiO$_3$ using DFT and the symmetry-adapted Landau-Ginzburg-Devonshire (LGD) framework. Our findings reveal that this coupling can induce double hysteresis at structural phase boundaries, independent of the FE single hysteresis. This behavior can be controlled through strain or layering within the perovskite structure, potentially expanding the range of materials with AFE-like properties and broadening their application possibilities. 

\begin{figure}
    \centering
    \includegraphics[width=0.85\linewidth]{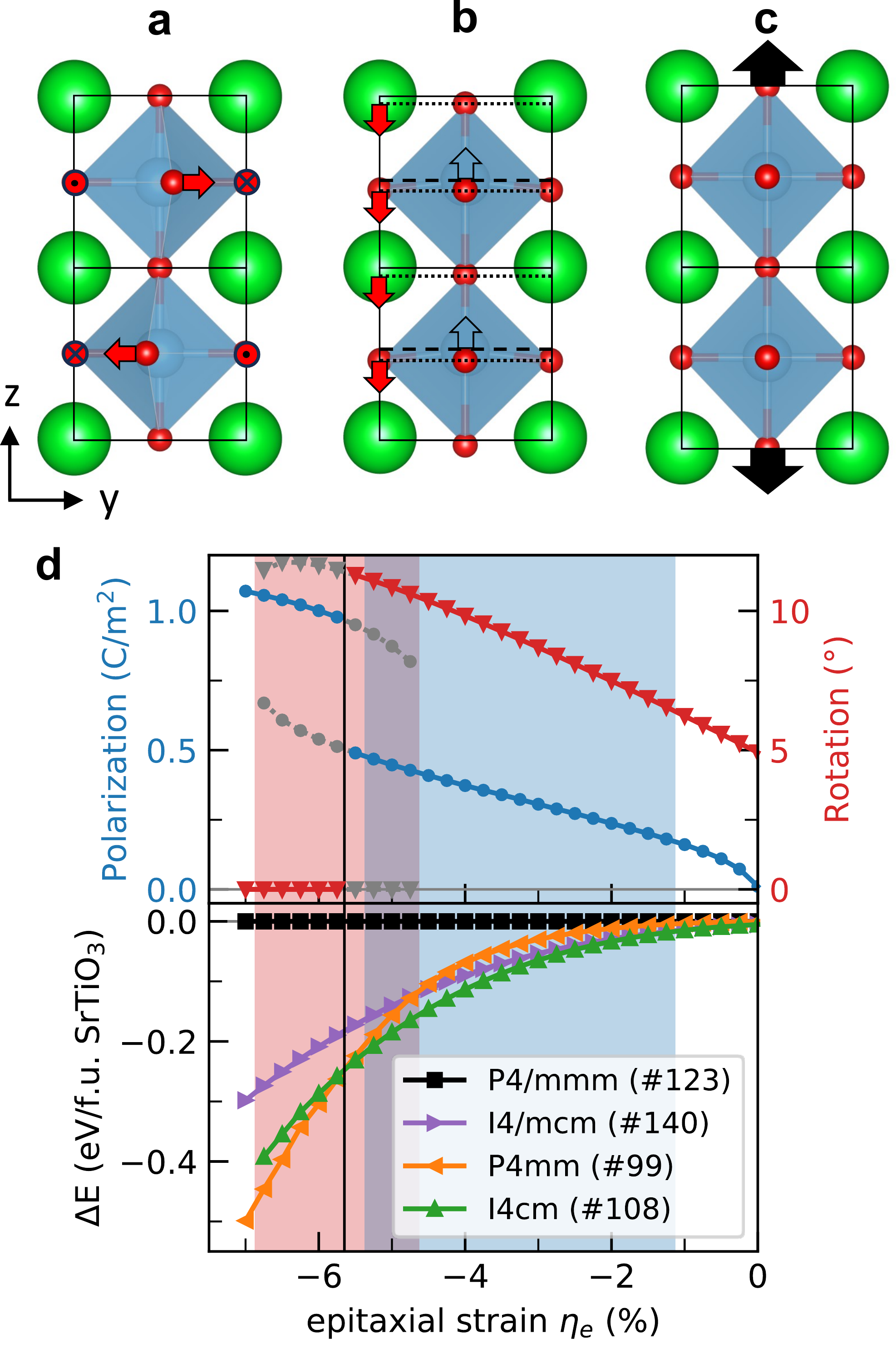}
    \caption{Strain-dependent ground state properties of SrTiO$_3$. (a) Depiction of the a$^0$a$^0$c$^-$ octahedral rotation, (b) polarization, and 
    (c) $\eta_{zz}$ strain in SrTiO$_3$. 
    (d) Upper panel: Spontaneous polarization (blue) and spontaneous octahedral rotation (red) in the ground state of SrTiO$_3$, with gray data points marking stable local minima. Lower panel: Relative energy for different phases (I4/mcm, P4mm, and I4cm) of SrTiO$_3$, referenced against the parent space group P4/mmm.
    }

    \label{strain}
\end{figure}

Fig.~\ref{strain}(a-c) illustrates the three order parameters associated with 
SrTiO$_3$ under compressive epitaxial strain: (a) the AFD rotation, (b) the polarization, and (c) the $\eta_{zz}$ strain. The AFD rotation, which corresponds to the R$_5^-$ irreducible representation (irrep) of the cubic space group, reduces the symmetry to the I4/mcm (\#140). Polarization, transforming as the $\Gamma_4^-$ irrep, lowers the symmetry further to the P4mm (\#99). When both polarization and AFD rotation are present, the symmetry is further reduced to I4cm (\#108). Finally, $\eta_{zz}$ strain lowers the symmetry to the P4/mmm (\#123) (see Table S1 for details).

Fig.~\ref{strain}(d) displays the spontaneous values of polarization and AFD rotation in the ground state (the state where the atomic positions and the $c$ axis length is optimized under biaxial strain boundary conditions) along with the relative energies of different  SrTiO$_3$ structures with space groups I4/mcm, P4mm, and I4cm, compared to the parent space group P4/mmm \cite{angsten2018epitaxial}. Without epitaxial strain, the DFT Kohn-Sham energies of the Pm$\bar{3}$m, I4/mcm, P4mm, and I4cm phases are comparable, within 6.0 meV/f.u. of each other. The spontaneous polarization converges to near zero at 0\%. Under larger compressive strain, all three phases become even lower in energy relative to P4/mmm.

The rate of change in energy with respect to strain varies significantly across different phases. I4cm, which hosts coexisting octahedral rotations and polarization, remains the lowest-energy phase for strain values up to --5.6\%. Beyond this strain threshold, the P4mm phase, which features only the polar structural distortion, becomes the lowest-energy configuration. Notably, for small strain values, the relative energy of P4mm decreases more slowly with strain compared to I4/mcm and I4cm, making it the least energetically favorable phase among the three. However, at larger compressive strain values, P4mm's relative energy decreases more sharply. As a result, the ground-state structure corresponds to the I4cm phase for epitaxial strains ranging from 0\% to --5.6\%, while for strains exceeding --5.6\%, the ground-state structure transitions to the P4mm phase.

As the ground state transitions from I4cm to P4mm under compressive strain, both spontaneous polarization and octahedral rotation angle exhibit a discontinuity at --5.6\% epitaxial strain. This is similar to reported case of EuTiO$_3$ \cite{yang2012understanding} and BaZrO$_3$ (Fig.~S3). In SrTiO$_3$, we observe distinct local minima of energy for both I4cm and P4mm phases between --4.8\% and --6.8\% strain, with data from these minima plotted as gray points in Fig.~\ref{strain}(d). The rapid decrease in energy for P4mm compared to I4cm, combined with the local minima that develop in highly strained SrTiO$_3$, indicates a strong coupling between polarization and octahedral rotation, leading to non-trivial hysteresis behavior that we discuss below.

\begin{figure*}
    \centering
    \includegraphics[width=1.0\linewidth]{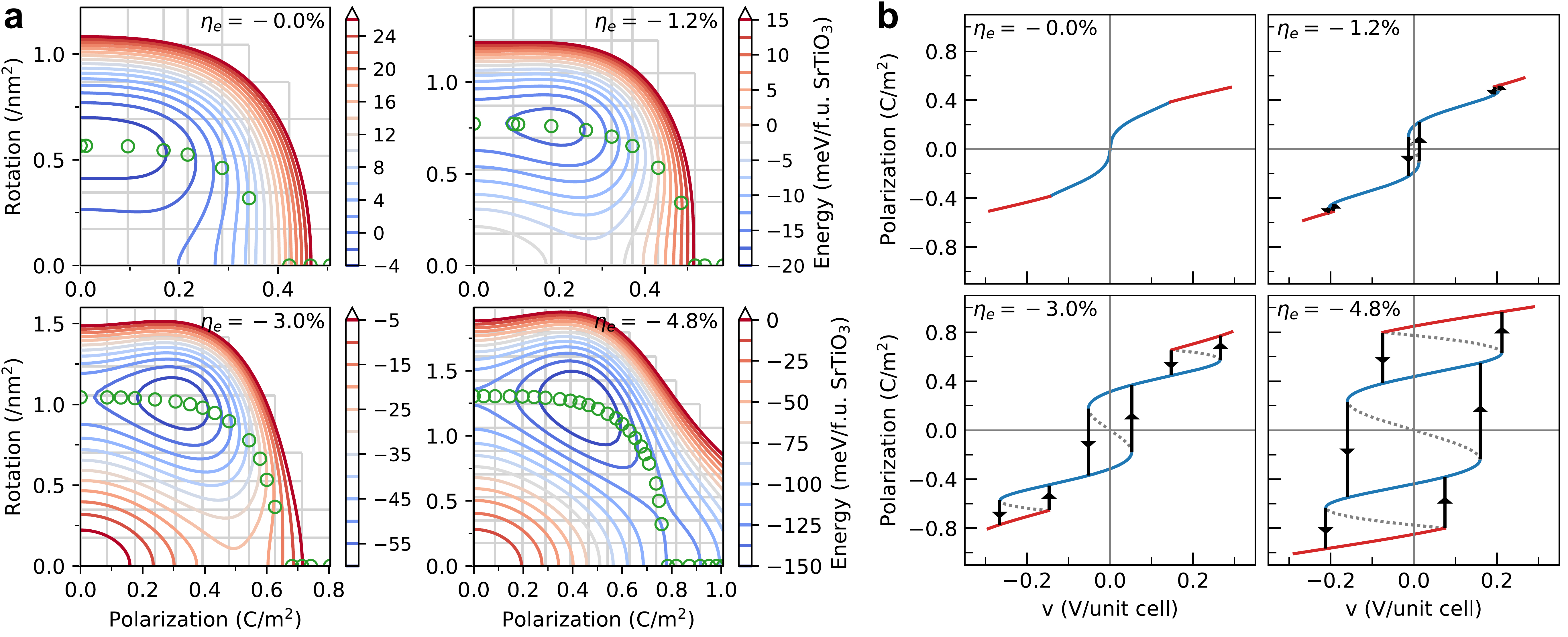}
    \caption{(a) The Kohn-Sham energy surface of SrTiO$_3$ as a function of $P$ and $Q$. 
    (b) Polarization-voltage ($P$-$v$) characteristics of SrTiO$_3$ under varying levels of epitaxial strain. An antiferroelectric (AFE) hysteresis loop emerges at the phase transition between the I4cm and P4mm phases at --1.2\% strain. As the strain increases, both ferroelectric and AFE hysteresis are amplified. At --4.8\% strain, the saddle point along the $P$ axis transitions into a local minimum. The blue, red, and dotted curves correspond to the stable I4cm phase, the stable P4mm phase, and the metastable I4cm phase, respectively.}
    \label{fig:regression}
\end{figure*}

Polarization transforms as the $\Gamma_4^-$ irrep, and it has 4 symmetry-adapted modes corresponding to the 4 different degrees of representing the displacements of Sr, Ti, and two types of O ions. For simplicity, we treat it as a single, one-dimensional order parameter with a displacement pattern that minimizes the electric enthalpy $H$ at a given value of polarization \cite{stengel2009electric, stengel2011band}\footnote{This approach differs from the common practice in first principles calculations, where the force constants eigenvector, which is kept fixed, is used to model the polarization. Our approach has the advantage of reproducing changes in the ratio of the displacements of different atoms due to higher order terms, while at the same time obeying the same, simplest form of the free energy.}. This approach reduces the dimensionality of the order parameter space without compromising the information along the minimum $H$ pathway. Additionally, it offers the advantage of linearizing the order parameter with a physically meaningful quantity that couples directly with the electric field. We predict the structure along the minimum $H$ pathway using the constrained-forces method, which was developed to simulate uniform electric fields in materials with high ionic contributions to permittivity \cite{sai2002theory, fu2003first, dieguez2006first, jung2023programmable}.

Using polarization $P$, octahedral rotation $Q$ (defined as the total displacements due to octahedral rotation divided by the initial volume without $\eta_{zz}$ strain), and strain $\eta_{zz}$, the following invariant polynomial expression of Kohn-Sham energy $E_{KS}$ can be employed to formulate a model describing the response to electric fields for materials that are materials that have a polarization only along the $z$-axis and have fixed epitaxial strain \cite{xue2015composition}.

\begin{align}
    \begin{split} \label{equ:Landau_6}
    E_{KS} &= \frac{a_2}{2}{P_z}^2 + \frac{b_2}{2}Q^2 \\
          &+ \frac{a_4}{4}{P_z}^4 + \frac{c_{22}}{2}{P_z}^2Q^2 + \frac{b_4}{4}Q^4 \\
          &+ \frac{a_6}{6}{P_z}^6 + \frac{c_{42}}{2}{P_z}^4Q^2 + \frac{c_{24}}{2}{P_z}^2Q^4 + \frac{b_6}{6}Q^6  \\
          &+ \frac{d_2}{2}{\eta_{zz}}^2 + e_{21}{P_z}^2\eta_{zz} + f_{21}Q^2{\eta_{zz}} 
    \end{split}
\end{align}

This simplified model only includes polarization along and octahedral rotation around a single axis ($z$), and hence it is relevant only under compressive biaxial strain applied on the $xy$ plane. We omit the subscripts in the rest of the paper. The energy expression includes combined terms of $P$ and $Q$ up to sixth order, strain terms up to second order, as well as the lowest order coupling terms between strain and other order parameters (electrostriction, which is second-order in $P$ and first order in strain; and `rotostriction' that is second order in $Q$ and linear in strain). $P$, $Q$, and $\eta$ are defined as independent of each other and as intensive quantities, which makes the coefficients in the free energy expansion scale with unit cell (u.c.) volume. This form of eq.~(\ref{equ:Landau_6}) is generic and is valid for many different $Q$'s representing various non-polar order parameters. Previously, the same equation was considered for AFEs where $Q$ represents the antipolar order parameter \cite{kittel1951theory,cross1967antiferroelectric,okada1969phenomenological,randall2021antiferroelectrics}.

We now define the electric enthalpy $H$ using reduced field variables to account for variable out-of-plane strain, $p_i = \Omega \bm{b}_i \cdot \bm{P}$ and $v_i = \bm{a}_i \cdot \bm{\mathcal{E}}$  where $\Omega$ is the u.c. volume, $\bm{a}_i$ are the direct lattice basis vectors, and $\bm{b}_i$ are the reciprocal lattice vectors (excluding the factor of $2\pi$). Defining $H = E_{KS} - p_i v_i$ and minimizing $H$ under fixed $v_i$ corresponds to imposing a boundary condition where a fixed electric potential bias $v_i$ is applied parallel to $\bm{a}_i$ across a u.c. of SrTiO$_3$ \cite{stengel2009electric}. Like polarization, we only consider voltage parallel to $\bm{a}_3$ along the $z$ axis.

This allows for the description of the state of SrTiO$_3$ at a given $v$, which satisfies $\partial H/\partial P = \partial H/\partial Q = \partial H/\partial \eta = 0$. From the last condition, we can express Kohn-Sham energy as a function of $P$ and $Q$ as:
\begin{align}
    \begin{split} \label{Landau_pq}
    E_{KS} &= \frac{a_2}{2}P^2 + \frac{b_2}{2}Q^2 + \left( \frac{a_4}{4} - \frac{{e_{21}}^2}{2c_{22}} \right)P^4 \\
          &+ \left( \frac{c_{22}}{2} - \frac{e_{21}f_{21}}{c_{22}} \right)P^2Q^2 + \left( \frac{b_4}{4} - \frac{{f_{21}}^2}{2c_{22}} \right)Q^4 \\
          &+ \frac{a_6}{6}P^6 + \frac{c_{42}}{2}P^4Q^2 + \frac{c_{24}}{2}P^2Q^4 + \frac{b_6}{6}Q^6 .
    \end{split}
\end{align}
Relaxation of the lattice constant $c$ in DFT under the constraints of fixed in-plane biaxial strain ($a$), at specific values of $P$ and $Q$ enables us to obtain the polynomial coefficients of eq.~(\ref{Landau_pq}) and plot $E_{KS}$ as a function of $P$ and $Q$.

In Fig.~\ref{fig:regression}(a), we present the energy from eq.~(\ref{Landau_pq}) with coefficients obtained by a fit to the Kohn-Sham energies obtained from DFT \footnote{Values of $P$ and $Q$ where DFT calculations were performed are shown as gray grid points in the background. The raw energy data from DFT, shown in the Fig. S1, shows identical behavior to Fig.~\ref{fig:regression}(a).}.
Points with different polarization along the minimum $H$ pathway are depicted as green hollow points. The ellipse-like pattern of AFD rotation in this pathway which diminishes after a critical polarization point, accurately reflects the model (Eq.~(S6)) and is consistent with the fixed displacement field calculations performed for SrTiO$_3$ \cite{hong2013electrically}. Unstable  $\Gamma_4^-$ and R$_5^-$ phonons are observed as negative curvature of energy along both $P$ and $Q$ axes near the origin in the plots, with the global minimum occurring at non-zero values of both $P$ and $Q$. As compressive strain increases, this global minimum shifts further from the $Q$ axis. Additionally, at --4.8\% epitaxial strain, a new local minimum emerges along the $P$ axis, corresponding to the P4mm phase.

In Fig.~\ref{fig:regression}(b), we show the polarization-voltage ($P$-$v$) plot of SrTiO$_3$ at various levels of epitaxial strain, obtained by minimizing the Landau polynomial under  fixed $v$. In the absence of epitaxial strain, the $P$-$v$ plot shows subtle ferroelectricity in the I4cm phase and no indication of double hysteresis at the I4cm-P4mm phase boundary. At --1.2\% epitaxial strain, double hysteresis emerges at the I4cm-P4mm boundary, accompanied by FE hysteresis in the P4mm phase (near zero field). In other words, a ``triple hysteresis'' behavior emerges \cite{pintilie2008coexistence,gao2024emergent}. Under larger compressive strain, for example at --3.0\% strain, all hysteresis loops expand and by --4.8\% strain, they overlap with each other. Notably, at this specific strain, the red line representing the P4mm phase crosses the $y$ axis, indicating that this phase is stable in the absence of the electric field. This corresponds to the formation of local minima in the P4mm phase and the possible stabilization of four spontaneous polarization values: two in the I4cm phase and two in the P4mm phase.

\begin{figure}
    \centering
    \includegraphics[width=0.8\linewidth]{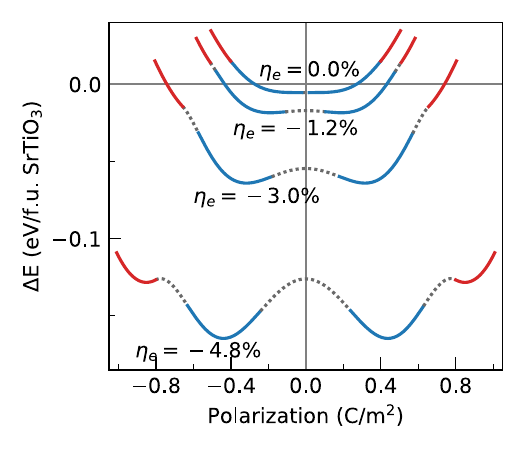}
    \caption{Energy-polarization plot of SrTiO$_3$ as derived from LGD theory under varying epitaxial strain values. The blue, red, and dotted curves represent the stable I4cm phase, stable P4mm phase, and metastable I4cm phase, respectively.}
    \label{pE}
\end{figure}

It is important to clarify that the triple hysteresis loops observed in Fig.~\ref{fig:regression}(b) result from two distinct phenomena: the FE single hysteresis loop (in the center) is within the I4cm phase, and involves a simple switching of polarization. The two loops that appear at larger field values are AFE-like double hysteresis loops between the I4cm and P4mm phases. These phenomena can also occur independently of each other. AFE double hysteresis \footnote{We propose two distinct scopes for the definition of antiferroelectricity. In a broad scope, antiferroelectricity encompasses all materials exhibiting double hysteresis behavior. In a narrower scope, antiferroelectricity is specifically attributed to materials that display double hysteresis behavior resulting from the coupling between an antipolar mode and a polar mode. We refer to materials that do not fit within this narrower definition as AFE-like.
} primarily arises from strong coupling between polar and non-polar order parameters, whereas the appearance of the FE hysteresis loop depends entirely on whether there is polar instability at the non-polar ground state structure. For SrTiO$_3$, the double hysteresis loops appear with polar instability, whereas for BaZrO$_3$, they appear without it (Fig. S5).

The energy-polarization plot in Fig.~\ref{pE} displays the progression of hysteresis behavior with increasing compressive biaxial strain. The blue, red, and dotted curves represent the stable I4cm, stable P4mm, and the metastable I4cm phases, respectively. The hysteresis behavior under applied strain is shown in Fig.~\ref{strain}(d) as the background shading color. Triple hysteresis loops are observed in the $P$-$v$ plot for the blue shaded strain range from --1.2\% to --4.6\%. In the purple shaded region, from --4.6\% to --5.4\%, triple hysteresis and four spontaneous polarization values are observed, indicated as a ``quadruple well'' in the energy-polarization plot. In the red shaded region beyond --5.4\% strain, the cascading triple hysteresis collapses into two hysteresis loops and ultimately into a single FE hysteresis loop that switches the polarization of the P4mm phase, at which octahedral rotations are completely suppressed by the polarization in its stable phases. 

\begin{figure*}
    \centering
    \includegraphics[width=0.85\linewidth]{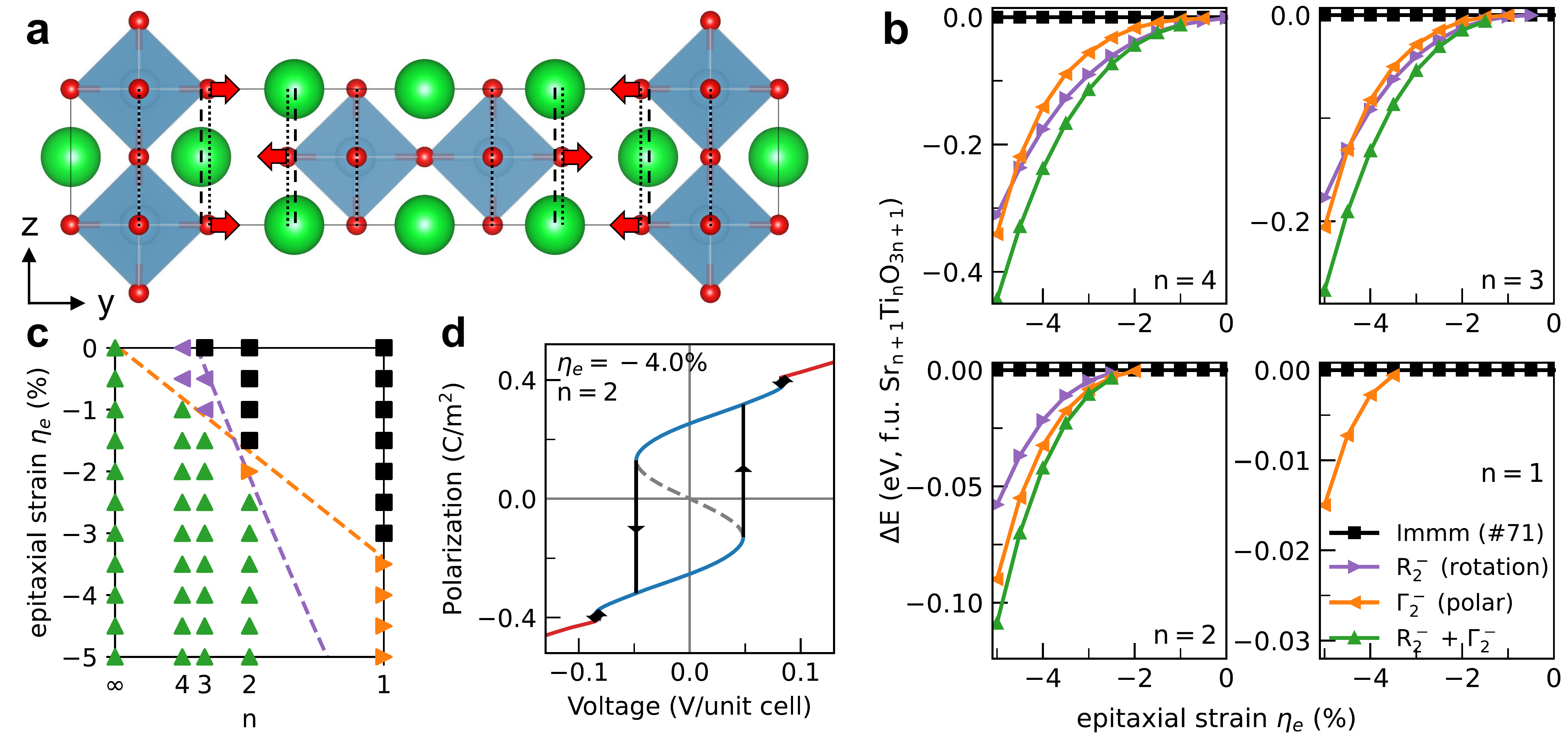}
    \caption{(a) Coherence-breaking rumpling motions in the 2-octahedral-layer Ruddlesden-Popper (RP) phase perovskite (Sr$_3$Ti$_2$O$_7$) hardens phonon modes associated with a$^0$a$^0$c$^-$ octahedral rotation (R$_2^-$) and polarization ($\Gamma_2^-$), offering additional control over the second-order coefficients $a_2$ and $b_2$. 
    (b) Strain phase diagram of Sr$_{n+1}$Ti$_n$O$_{3n+1}$ for $n$ values from 1 to 4. 
    (c) Most stable phase of Sr$_{n+1}$Ti$_n$O$_{3n+1}$ as a function of epitaxial strain and $n$. 
    (d) Polarization-voltage hysteresis plot of Sr$_3$Ti$_2$O$_7$ at --4\% strain.}

    \label{RP}
\end{figure*}

From Fig.~\ref{fig:regression}(b), it is evident that the AFE hysteresis is associated with a sign change in $\dd v / \dd P$ at the phase boundary. The conditions for this sign change, derived from the simplified eq.~(\ref{equ:Landau_6}) up to fourth order in $P$ and $Q$, are given by (Eq.~S13): 
\begin{gather} \label{sign_change} -3\frac{a_4}{c_{22}} < \frac{a_2}{-b_2} < -3\frac{a_4}{c_{22}} + 2\frac{c_{22}}{b_4} \end{gather}
In pretty much every known perovskite, compressive epitaxial strain strongly affects both the R$_5^-$ and $\Gamma_4^-$ phonon frequencies, which directly affects the coefficients $a_2$ and $b_2$ \cite{antons2005tunability,saha2024effect}. In SrTiO$_3$, the net effect of compressive strain on ${a_2}/{-b_2}$ is negative, leading to the appearance of AFE hysteresis with increasing compressive strain.

Although SrTiO$_3$ is extensively studied, there are no reports of double hysteresis in thin-film SrTiO$_3$ to the best of our knowledge. Both strain and temperature are critical factors for observing double hysteresis, and only a narrow range of these variables is expected to exhibit the phenomenon. Additionally, even though the onset of the triple hysteresis behavior doesn't require an experimentally unfeasible amount of strain, it is likely that a prohibitively high electric field is likely required to observe the transition between the two polar phases with and without octahedral rotations \cite{gao2023determines}. High-$\kappa$ SrTiO$_3$ suffers from a low breakdown field of $\sim$1-2 MV/cm \cite{mcpherson2002proposed}. 
Fig.~\ref{fig:regression}(b) shows an estimate of the field required by assuming coherent transition (which typically overestimate the field requirement of switching via domain-wall motion \cite{beckman2009ideal}) at --1.2\% epitaxial strain to be 5.1 MV/cm, about 15 times larger than the FE switching field. The simplified expression for the critical AFE-to-FE transition field $\mathcal{E}_{\mathrm{AFE}}$ under the same conditions as eq.~(\ref{sign_change}) is (Eq.~(S17)): 
\begin{equation} \label{switching_field}
    |\mathcal{E}_{\mathrm{AFE}}| = \frac{2}{\Omega b_4\sqrt{{c_{22}}^2-a_4b_4}} \left( \frac{a_2b_4-b_2c_{22}}{3} \right)^{\frac{3}{2}}
\end{equation}
Equations \ref{sign_change} and \ref{switching_field} highlight the desired material properties for AFE applications, which include strong coupling (high $c_{22}$), ferroelectricity or high-$\kappa$ (low $a_2$), low fourth-order parameters (low $a_4$ and $b_4$), and weak (but non-zero) non-polar structural distortions (low $-b_2$). Specifically, low $-b_2$ and $a_2$ are crucial for reducing $|\mathcal{E}_{AFE}|$.

While epitaxial strain provides an effective means to control the  second-order parameters to induce AFE-like double hysteresis, it is but a single parameter, and hence can only scan a one-dimensional parameter space. To achieve two-dimensional, more comprehensive control over second-order parameters and further reduce the critical AFE-to-FE transition field, additional strategies are needed to manipulate phonon frequencies. One promising approach is to explore the Ruddlesden-Popper (RP) phases of perovskite structures, which is already extensively used to control the emergence of ferroelectricity, as well as other parameters such as loss \cite{birol2011interface, lee2013exploiting}. Fig.~\ref{RP}(a) represent the antipolar rumpling observed in RP phases known to disrupt layer coherence and weaken both AFD rotational and polar instabilities \cite{lee2010structural,birol2011interface}.

For compressive biaxial strain on the $xz$-plane in Fig.~\ref{RP}(a), the out-of-plane ($y$) polarization is significantly suppressed. For this reason, we consider $xy$-biaxial strain with polarization and rotation along the $z$ axis \footnote{While this is not the commonly used growth direction in these systems using methods such as molecular beam epitaxy, it is in principle possible if the right substrates are identified for epitaxial growth, or with the help of the new generation mechanical strain cells \cite{najev2022uniaxial}.}. The strain phase diagram for Sr$_{n+1}$Ti$_n$O$_{3n+1}$, with the number of octahedral layers $n$ ranging from one to four, is shown in Fig.~\ref{RP}(b) and (c). Reducing the number of octahedral layers in each RP block enhances the effect of antipolar rumpling ($\Gamma_1^+$) and hardening of the AFD rotational ($\mathrm{R}_2^-$) and polar ($\Gamma_2^-$) phonons. Between these two, reducing the number of octahedral layers hardens the polar phonon more than the AFD rotational phonon. According to eq.~(\ref{sign_change}) and (\ref{switching_field}), assuming that higher-order coefficients including the biquadratic coupling coefficient are not significantly affected, this suggests that an AFE-like double hysteresis can form with reduced AFD rotation ($-b_2$), which corresponds to a lower critical AFE-to-FE field $|\mathcal{E}_{\mathrm{AFE}}|$. A small number of octahedral layers in the RP phase, combined with high biaxial strain, is likely to result in a reduced $|\mathcal{E}_{\mathrm{AFE}}|$.

To demonstrate this, we show the polarization-voltage plot for the two-octahedral-layer RP phase Sr$_3$Ti$_2$O$_7$ under --4.00\% epitaxial strain in Fig.~\ref{RP}(d). The AFE hysteresis emerges with a switching voltage reduced to as low as 0.084 V/u.c. (2.1 MV/cm), compared to 0.205 V/u.c. (5.1 MV/cm) in bulk SrTiO$3$ at --1.2\% epitaxial strain. This strategic control of phonon frequencies using two parameters -- both strain and oxide layering -- reduces $|\mathcal{E}{_\mathrm{AFE}}|$ by almost 60\% compared to the simpler one-parameter control using epitaxial strain alone.

In conclusion, we present the discovery of AFE-like double hysteresis without antipolar structural features, enabled by octahedral rotations in strained perovskites and its RP phases. Additionally, we propose strategies for designing materials with low critical AFE-to-FE fields. Our results suggest a possible novel mechanism of experimentally observed antiferroelectricity, and show potential expansion in space of candidate AFEs. Also, this work highlights the importance of symmetry-mode couplings and need for a deeper understanding of higher order terms in the Landau theory to advance the discovery and design of AFEs. 

This work is supported by the  Office of Naval Research Grants N00014-20-1-2361 and N00014-24-1-2082. We also acknowledge the Minnesota Supercomputing Institute (MSI) at the University of Minnesota for providing computational resources that contributed to this research.

\end{document}


{\let\newpage\relax\maketitle}

\section*{Methods}
Periodic density functional theory (DFT) calculations were performed with the a customized version of Vienna ab initio simulation package (VASP) 6.4.1 \cite{kresse1996efficient} modified for constrained-forces calculations \cite{sai2002theory,fu2003first,jung2023programmable,seongjoo_jung_2024_11554391}. No spin-polarization was observed in the systems studied. A generalized gradient approximation (GGA) exchange-correlation functional PBEsol was used in all calculations \cite{perdew2008restoring}. The projector augmented wave (PAW) method to describe atom cores and the plane wave basis set was expanded to a kinetic energy maximum of 520 eV for Kohn-Sham orbitals.   Tetrahedron smearing with Bl\"{o}chl corrections imposed on electrons at the Fermi level. VESTA software \cite{momma2011vesta} was used for building and visualizing crystal structures.

6 $\times$ 6 $\times$ 5 $\Gamma$-centered \textbf{k}-point mesh was used to sample the Brillouin zone of the 20-atom supercell of perovskite ABO$_3$. The constrained-forces method were used to estimate polarized structures of ABO$_3$. Optimization of structures was converged below 10$^{-7}$ eV of electronic energy changes and 1.0 meV/\AA~of maximum ionic force differences. Lattice parameters for cubic SrTiO$_3$ and BaZr$O_3$ (Pm$\bar{3}$m, \#221) were 3.896 \AA~and 4.191 \AA. Polarization was calculated using the Berry phase formalism of the modern theory of polarization \cite{king1993theory}. 

The order parameter amplitudes were defined as the sum of the displacements of the ions, normalized by the volume, ensuring compatibility with the use of polarization as an order parameter:
\begin{gather}
    Q=\frac{\Sigma {|u_{i\alpha}|}}{\Omega_0}
\end{gather}
Here, $u$ represents the displacement of an ion based on the lattice parameters of the high-symmetry (P4/mmm or Immm) structure without strain, and $\Omega_0$ denotes the volume of this high-symmetry structure. Both indices are summed over: The index $i$ iterates over all ions in the unit cell, while $\alpha$ iterates over each Cartesian directions. This definition scales linearly with the more commonly used order parameter definition when the order parameter involves only a single degree of freedom \cite{cowley1980structural}:
\begin{gather}
    Q=\sqrt{\frac{\Sigma {u_{i\alpha}}^2}{N}}
\end{gather}
where $N$ is the number of primitive cells in the unit cell. The same expression without normalization factor $\sqrt{N}$ represents the mode amplitude.

For Kohn-Sham energy regression, $z$ internal coordinates were fixed for structures with same polarization obtained using constrained-forces method. a$^0$a$^0$c$^-$ octahedral rotation was kept constant by fixing $x$ and $y$ internal coordinates to specified values. Space group representations, displacements and mode amplitudes were referred from Bilbao 
Crystallographic Server \cite{aroyo2011crystallography} and calculated using ISODISTORT from the ISOTROPY software suite \cite{campbell2006isodisplace,stokes_isodisplace}, and the Eq. (3) from the main text were used for regression for BaZrO$_3$ and SrTiO$_3$ with epitaxial compressive strain values lower than -4.75 \%. For SrTiO$_3$ regression with more compressive epitaxial strain, 8th order $P^nQ^m$ terms ($\frac{a_8}{8}P^8$, $\frac{c_{62}}{2}P^6Q^2$, $\frac{c_{44}}{4}P^4Q^4$, $\frac{c_{26}}{2}P^2Q^6$, $\frac{b_8}{8}Q^8$) were added to energy expression to accommodate more extrema of $E_{KS}(P)$ function.

5 $\times$ 1 $\times$ 5 $\Gamma$-centered \textbf{k}-point mesh was used to sample the Brillouin zone of the 48-atom supercell of Sr$_3$Ti$_2$O$_7$ Ruddlesden-Popper (RP) phase. 5 $\times$ 5 $\times$ 1 $\Gamma$-centered \textbf{k}-point mesh was used to sample the Brillouin zone of [SrTiO$_3$]$_n$-[BaZrO$_3$]$_n$ structures. Optimization of structures was converged below 10$^{-6}$ eV of electronic energy changes and 1.0 meV/\AA~of maximum ionic forces. Shear components of strain were ignored for optimization of the structures. (The perovskite structures considered in this study are orthorhombic or higher in symmetry, and hence are not affected from this approximation. It can only affect the C2/m and Cm phase of the Ruddlesden-Popper structure considered.) Lattice constants $a$ and $c$ for highest symmetry Sr$_3$Ti$_2$O$_7$ RP structure (I4/mmm, \#139) were 3.890 \AA~and 10.127 \AA. 

\newpage
\section*{Representations of structural distortions from different parent phases}
\begin{table}[h]
	\centering
	\caption{Irreducible representations (irreps) of the relevant structural distortions, i.e. the order parameters entering the Landau expansion, derived from different parent structures. While biaxial strain imposes a tetragonal P4/mmm parent phase, we reference irreducible representations from the cubic phase to ensure direct alignment with existing literature. Note that the strain tensor component $\eta_{zz}$ belongs to the reducible representation $\Gamma_{1}^+ \oplus \Gamma_{3}^+$. 
    Additionally, the irreps from the Pm$\bar{3}$m phase are three-dimensional, whereas those from the P4/mmm phase are one-dimensional. As a result, our energy expansion does not require in-plane components of polarization and rotation.}
	\label{sup_table} 

        \renewcommand{\arraystretch}{1.5}
	\begin{tabular}{ccc} 
		\\
		\hline
		  \multirow{2}{3.0cm}{Order Parameter} & \multicolumn{2}{c}{Irrep} \\
             & Pm$\bar{3}$m & P4/mmm\\
		\hline
            $\eta_{zz}$ & $\Gamma_{1}^+ \oplus \Gamma_{3}^+$ & $\Gamma_{1}^+$  \\
		$P$ & $\Gamma_{4}^-$ & $\Gamma_{3}^-$ \\
		$Q$ & $\mathrm{R}_{5}^-$ & $\mathrm{A}_{4}^-$ \\
		\hline
	\end{tabular}
\end{table}

\clearpage
\section*{Derivation of double hysteresis from Landau equation}
\setcounter{section}{1}

We start from simpler version of Landau energy expression compared to eq.~(1) in the main text, only using terms up to 4th order $P$ and $Q$ where only polarization, rotation and electric field along $z$ axis perpendicular to strained plane are considered. 
\begin{gather} 
    E_{KS} = \frac{a_2}{2}P^2 + \frac{a_4}{4}P^4 + \frac{b_2}{2}Q^2 + \frac{b_4}{4}Q^4 + \frac{c_{22}}{2}P^2Q^2 \label{Landau_4} \\
    H = E_{KS} - APv
\end{gather}
 $A$ is the cross-sectional area of the strained surface from the unit cell. We restrict highest order coefficients $a_4, b_4 > 0$ to prevent divergence. We will also restrict $c_{22} > 0$ for the scope of this paper. 

Solution to this equation is when $P$ and $Q$ satisfy the following equations.
\begin{align}
    \frac{1}{\Omega} \pdv{H}{P} &= a_2P + a_4P^3 + c_{22}PQ^2 - Av = 0 \label{dEdP} \\ 
    \frac{1}{\Omega} \pdv{H}{Q} &= b_2Q + b_4Q^3 + c_{22}P^2Q = 0 \label{dEdQ} \\
    \frac{1}{\Omega} \pdv[{2}]{H}{Q} &= b_2 + 3b_4Q^2 + c_{22}P^2 \geq 0 \label{ddEdQQ}
\end{align}
We put no restriction in terms of second partial derivative with respect to $P$. Even though the points with positive $\partial^2H/\partial P^2$ are not local minima, they are necessary for expression of energy as a function of $P$ and to explain the path of polarization switching. 

The relations of $P$ and $Q$ of the solutions can be explained from eq.~(\ref{dEdQ}). for positive $b_2$, the only solutions satisfying eq.~(\ref{dEdQ}) is trivial case of $Q=0$. If $a_2 > 0$, it corresponds to simple case of dielectric material. If $a_2 < 0$, it corresponds to simple ferroelectric material with no other structural distortions. As there is no phase transition that derives from a non-polar mode, double hysteresis cannot occur for positive $b_2$. In other words, a structural instability is necessary, but not a sufficient factor for double hysteresis. For negative $b_2$, from eq.~(\ref{dEdQ}) and (\ref{ddEdQQ}):
\begin{equation} \label{Q(P)}
    Q(P) = \begin{cases}
        \sqrt{\frac{-b_2 - c_{22}P^2}{b_4}} &\text{$\left( |P| < \sqrt{-\frac{b_2}{c_{22}}} \right)$} \\
        0 &\text{$\left( |P| \geq \sqrt{-\frac{b_2}{c_{22}}} \right)$}
    \end{cases}
\end{equation}

Note that the point of $P$ where $Q$ is suppressed, the phase transition point $(P,Q) = (\sqrt{-b_2/c_{22}},0)$ does not coincide with the minimum along $P$ axis when $a_2 < 0$ at $v = 0$, $(P,Q) = (\sqrt{-a_2/a_4},0)$. From eq.~(\ref{Q(P)}), we find the origin of double hysteresis, which is the discontinuity of $\dd Q/\dd P$ at phase boundary.
\begin{equation} \label{dQdP}
    \dv{Q}{P} = \begin{cases}
        \frac{-c_{22}P}{b_4Q} &\text{$\left( |P| < \sqrt{-\frac{b_2}{c_{22}}} \right)$} \\
        0 &\text{$\left( |P| > \sqrt{-\frac{b_2}{c_{22}}} \right)$}
    \end{cases}
\end{equation}

Now we will calculate the range of parameters that forms hysteresis at the phase boundary.  From eq.~(\ref{dEdP}) and (\ref{dEdQ}),
\begin{gather}
    Av = \pdv{E_{KS}}{P} = \dv{E_{KS}}{P}\\
    A\dv{v}{P} = \dv[2]{E_{KS}}{P} = \left( \pdv[2]{E_{KS}}{P} + \pdv{E_{KS}}{P}{Q} \dv{Q}{P} \right) \label{dEfdP}
\end{gather}
Using eq.~(\ref{dQdP}):

\begin{equation} \label{dvdP}
    A\dv{v}{P} = \begin{cases}
          a_2 +  c_{22}Q^2 + \left(3a_4 - \frac{2{c_{22}}^2}{b_4}\right) P^2   &\text{$\left( |P| < \sqrt{-\frac{b_2}{c_{22}}} \right)$} \\
          a_2 + 3a_4P^2 &\text{$\left( |P| > \sqrt{-\frac{b_2}{c_{22}}} \right)$}
    \end{cases}
\end{equation}

Conditions of hysteresis forming is when $\dd P/\dd v$ changes sign at the phase boundary $(P,Q) = (\sqrt{-b_2/c_{22}},0)$ 
\begin{gather} \label{cond1}
    a_2 + -3\frac{a_4b_2}{c_{22}} + 2\frac{b_2c_{22}}{b_4} < 0 \\
    a_2 + -3\frac{a_4b_2}{c_{22}} > 0
\end{gather}
which combines to eq.~(4) in the main text:
\begin{gather}
    -3\frac{a_4}{c_{22}} < \frac{a_2}{-b_2} < -3\frac{a_4}{c_{22}} + 2\frac{c_{22}}{b_4}
\end{gather}

There is an additional condition, which restricts entire $\dd P/\dd v$ to remain negative where $|P| < \sqrt{-\frac{b_2}{c_{22}}}$
\begin{gather} 
    \left. A\dv{v}{P} \right|_{P=0} = a_2 - \frac{b_2c_{22}}{b_4} > 0
\end{gather}
Which rearranges to:
\begin{gather} \label{cond3}
    -\frac{c_{22}}{b_4} < \frac{a_2}{-b_2}
\end{gather}

Antiferroelectric (AFE) transition voltage ($v_{AFE}$) corresponds to the voltage at $P$ where $\dd v/\dd P = 0$. From eq.~(\ref{Q(P)}) and (\ref{dvdP}):
\begin{gather} \label{P_AFE}
    \left| P \right|_{\dd v/\dd P=0} = \sqrt{\frac{{c_{22}}^2-a_4b_4}{a_2b_4-b_2c_{22}}}
\end{gather}
Note that the denominator of eq.~(\ref{P_AFE}) is positive from eq.~(\ref{cond3}), and that the numerator is also positive form eq.~(\ref{cond1}) and (\ref{cond3}). Using eq.~(\ref{P_AFE}) and (\ref{dEdP}), the equation for $v_{AFE}$ and $\mathcal{E}_{AFE}$ in the main text (eq. 5) is obtained.

\begin{equation} 
    A|v_{AFE}| = \frac{2}{b_4\sqrt{c_{22}^2-a_4b_4}} \left( \frac{a_2b_4-b_2c_{22}}{3} \right)^{\frac{3}{2}}
\end{equation}

\section*{DFT results of SrTiO$_3$}

\begin{figure}[H]
    \centering
    \includegraphics[width=0.7\linewidth]{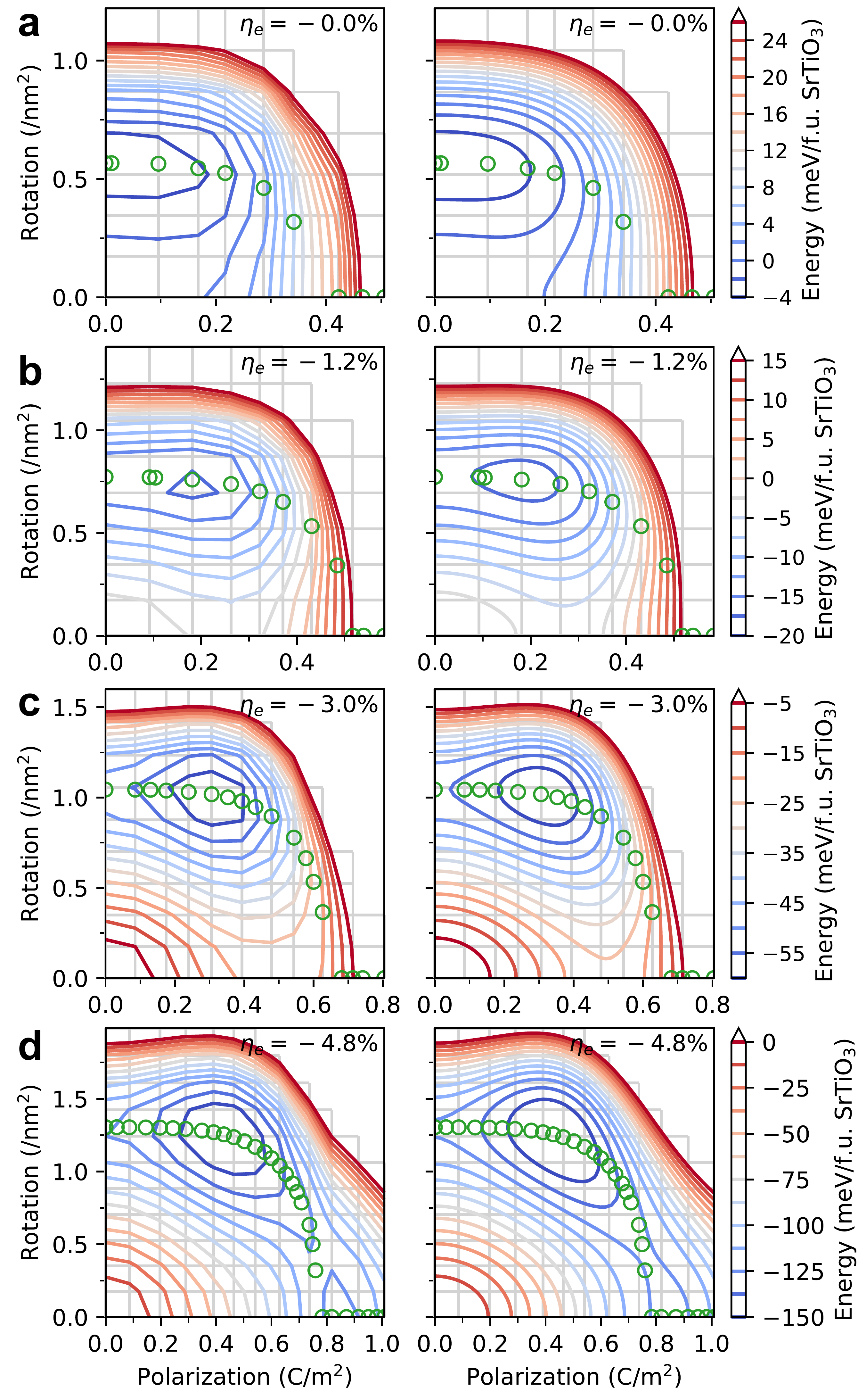}
    \caption{Comparison of DFT energy contours (left) and Landau-Devonshire energy contours using the coefficients regressed from DFT (right) of SrTiO$_3$ at different values of epitaxial strain. (a) 0.0 \% (b) --1.2 \% (c) --3.0 \% (d) --4.8 \%
    }
\end{figure}

\begin{figure}[H]
    \centering
    \includegraphics[width=0.7\linewidth]{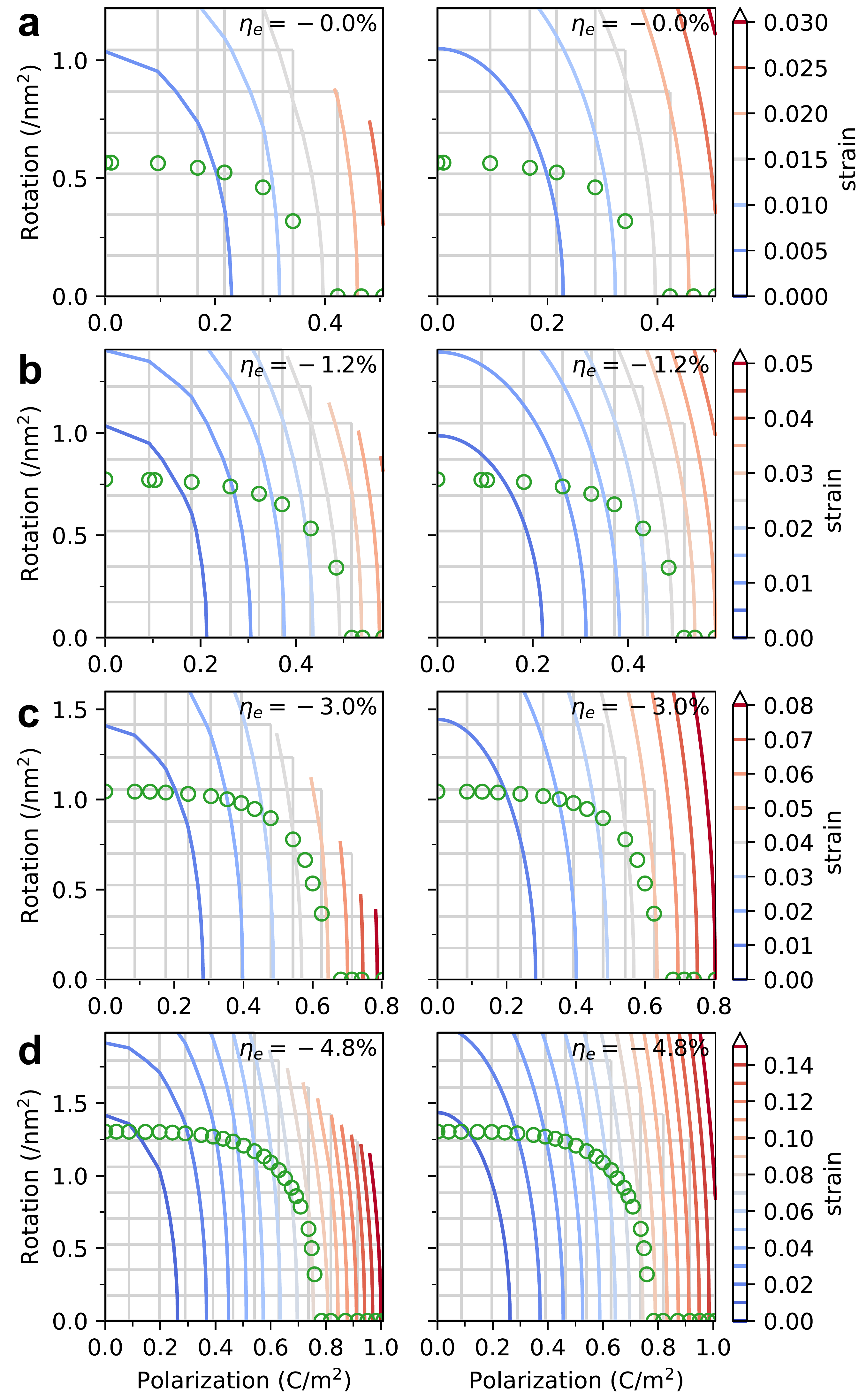}
    \caption{Comparison of DFT strain contours (left) and strain contours obtained from Landau-Devonshire energy expression using the coefficients regressed from DFT (right) of SrTiO$_3$ at different values of epitaxial strain. (a) 0.0 \% ($c_0=3.896$ \AA) (b) --1.2 \% ($c_0=3.926$ \AA) (c) --3.0 \% ($c_0=3.971$ \AA) (d) --4.8 \% ($c_0=4.021$ \AA)
    }
\end{figure}

\newpage
\section*{Double hysteresis of BaZrO$_3$}
\begin{figure}[h]
    \centering
    \includegraphics[width=0.5\linewidth]{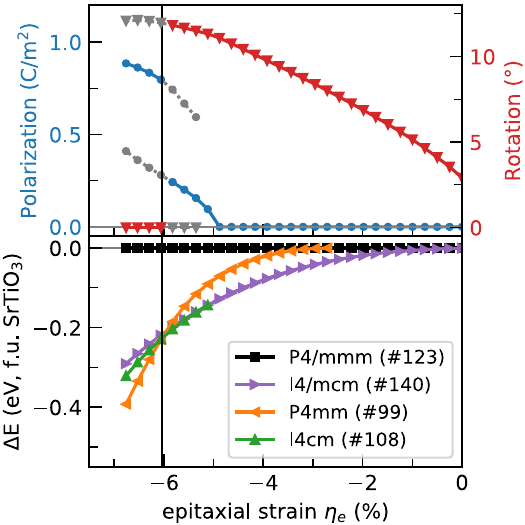}
    \caption{Strain-dependent ground state properties of BaZrO$_3$. Upper panel: Spontaneous polarization (blue) and spontaneous octahedral rotation (red) in the ground state of BaZrO$_3$, with gray data points indicating stable local minima. Lower panel: Relative energy per formula unit of difference phases (I4/mcm, P4mm, and I4cm) of BaZrO$_3$, referenced to the parent space group P4/mmm. Like SrTiO$_3$ and EuTiO$_3$, strong discontinuity in both spontaneous polarization and rotation during phase transition from I4/mcm to P4mm indicate strong coupling between rotation and polarization.
    }
    \label{sup_BZO_strain}
\end{figure}

\begin{figure}[H]
    \centering
    \includegraphics[width=0.65\linewidth]{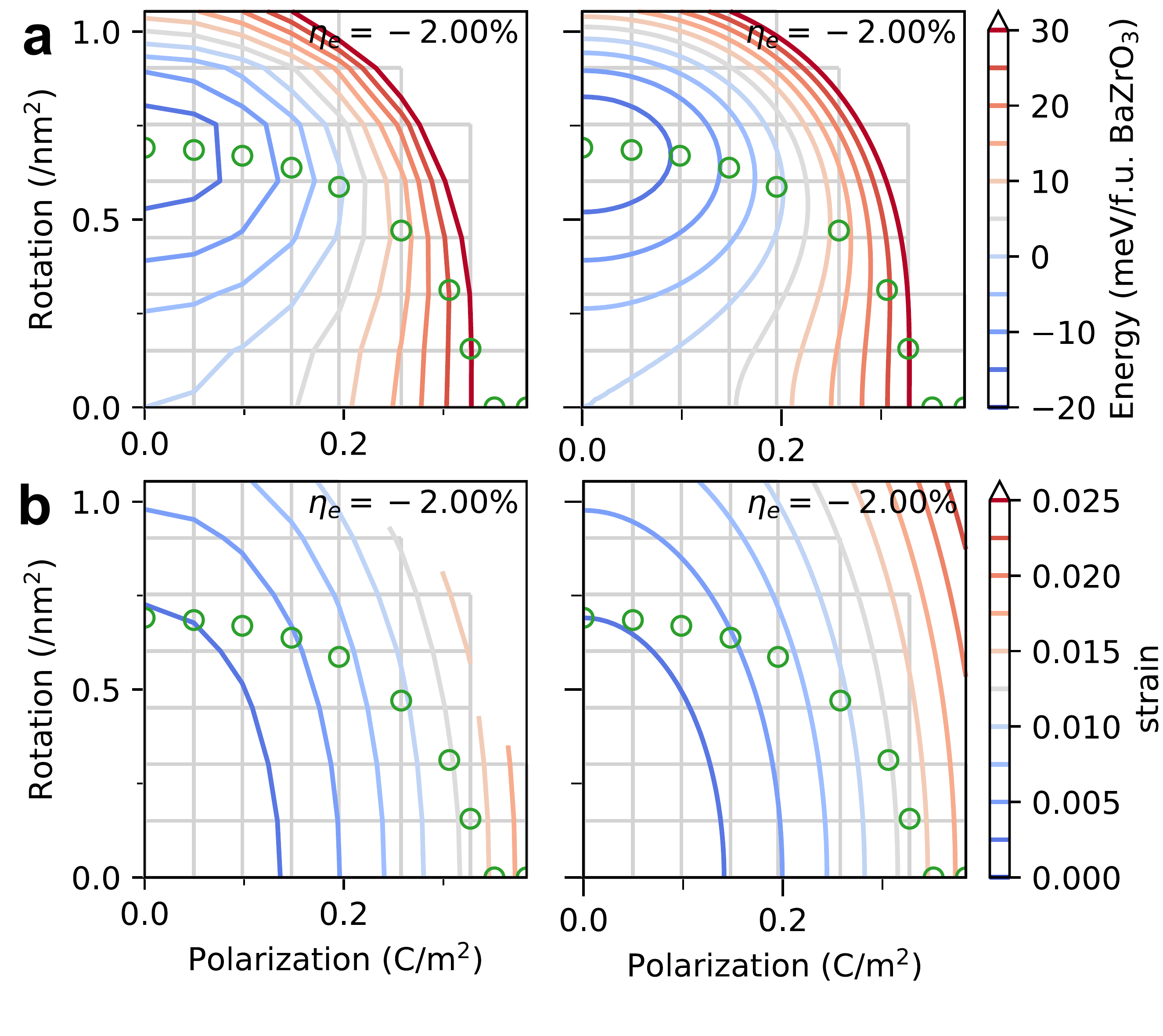}
    \caption{(a) Comparison of DFT energy contours (left) and Landau-Devonshire energy contours using the coefficients regressed from DFT (right) of BaZrO$_3$ at --2.0 \% epitaxial strain. (b) Comparison of DFT strain contours (left) and strain contours obtained from Landau-Devonshire energy expression using the coefficients regressed from DFT (right) of BaZrO$_3$ at --2.0 \% epitaxial strain ($c_0=4.236$ \AA).
    }
\end{figure}

\begin{figure}[H]
    \centering
    \includegraphics[width=0.4\linewidth]{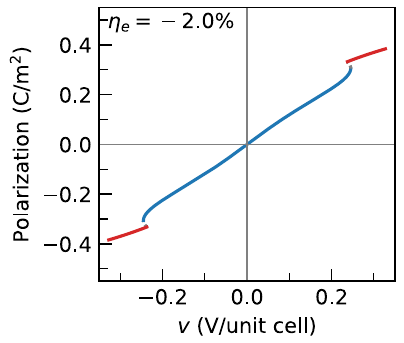}
    \caption{Polarization-voltage plot for BaZrO$_3$ at --2.0 \% epitaxial strain.
    }
\end{figure}

\bibliographystyle{unsrt}